\documentclass[showpacs,10pt,twocolumn,prb]{revtex4-1}

\usepackage{amsmath}
\usepackage{amssymb}
\usepackage{graphicx}
\usepackage{amssymb}
\usepackage{graphics}
\usepackage{epsfig}
\usepackage{CJK}
\usepackage{color}

\begin{document}

\begin{CJK*}{GBK}{Song}
\title{Critical behavior and magnetocaloric effect in VI$_3$}
\author{Yu Liu,$^{1}$ Milinda Abeykoon,$^{2}$ and C. Petrovic$^{1}$}
\affiliation{$^{1}$Condensed Matter Physics and Materials Science Department, Brookhaven National Laboratory, Upton, New York 11973, USA\\
$^{2}$National Synchrotron Light Source II, Brookhaven National Laboratory, Upton, New York 11973, USA}
\date{\today}

\begin{abstract}
Layered van der Waals ferromagnets are promising candidates for designing new spintronic devices. Here we investigated the critical properties and magnetocaloric effect connected with ferromagnetic transition in layered van der Waals VI$_3$ single crystals. The critical exponents $\beta = 0.244(5)$ with a critical temperature $T_c = 50.10(2)$ K and $\gamma = 1.028(12)$ with $T_c = 49.97(5)$ K are obtained from the modified Arrott plot, whereas $\delta = 5.24(2)$ is obtained from a critical isotherm analysis at $T_c = 50$ K. The magnetic entropy change $-\Delta S_M(T,H)$ features a maximum at $T_c$, i.e., $-\Delta S_M^{max} \sim$ 2.64 (2.27) J kg$^{-1}$ K$^{-1}$ with out-of-plane (in-plane) field change of 5 T. This is consistent with $-\Delta S_M^{max}$ $\sim$ 2.80 J kg$^{-1}$ K$^{-1}$ deduced from heat capacity and the corresponding adiabatic temperature change $\Delta T_{ad}$ $\sim$ 0.96 K with out-of-plane field change of 5 T. The critical analysis suggests that the ferromagnetic phase transition in VI$_3$ is situated close to a three- to two-dimensional critical point. The rescaled $\Delta S_M(T,H)$ curves collapse onto a universal curve, confirming a second-order type of the magnetic transition and reliability of the obtained critical exponents.
\end{abstract}
\maketitle
\end{CJK*}

\section{INTRODUCTION}

Layered intrinsically ferromagnetic (FM) semiconductors hold great promise for both fundamental physics and applications in spintronic devices.\cite{McGuire0, McGuire, Huang, Gong, Seyler} CrI$_3$ has recently attracted much attention since the long-range magnetism persists in monolayer with $T_c$ of 45 K.\cite{Huang} Intriguingly, the magnetism in CrI$_3$ is layer-dependent, from FM in monolayer, to antiferromagnetic (AFM) in bilayer, and back to FM in trilayer.\cite{Huang} In van der Waals (vdW) heterostructures formed by an ultrathin CrI$_3$ and a monolayer WSe$_2$, the WSe$_2$ photoluminescence intensity strongly depends on the relative alignment between photoexcited spins in WSe$_2$ and the CrI$_3$ magnetization.\cite{Zhong} The magnetism in ultrathin CrI$_3$ could also be controlled by electrostatic doping, which provides great opportunities for designing magneto-optoelectronic devices.\cite{Jiang, Huang1} Very recently, the two-dimensional (2D) ferromagnetism has also been predicted in VI$_3$ monolayer with a calculated T$_c$ of 98 K, higher than that in CrI$_3$.\cite{He}

Bulk CrI$_3$ and VI$_3$ belong to a well-known family of transition metal trihalides MX$_3$ (X = Cl, Br and I).\cite{Juza,Dillon} When compared to CrI$_3$, in which the chromium has a half filled t$_{2g}$ level yielding S = 3/2, the vanadium in VI$_3$ has two valence electrons that half fill two of the three degenerate t$_{2g}$ states yielding S = 1.\cite{Son,Kong,Tian} Bulk VI$_3$ is an insulating 2D ferromagnet with $T_c$ = 55 K and crystallizes in a layered structure.\cite{Trotter,Handy,Wilson} Each V ion is centered in an octahedron of I ions, form a honeycomb lattice within the $ab$ plane [inset in Fig. 1(a)], similar with CrI$_3$. There is a structural transition at $\sim$ 80 K above $T_c$, however, the detailed symmetry of the high- or low-temperature structure is still not settled. Tian et al. describes analysis of single crystal x-ray diffraction (XRD) data and concludes that the high temperature structure is monoclinic, and the low temperature structure is trigonal,\cite{Tian} while Son et al. describes powder XRD and arrives at the inverse conclusion,\cite{Son} calling for further study. Density functional theory (DFT) calculations suggest that the VI$_3$ not only hosts the long-range ferromagnetism down to a monolayer but also exhibits Dirac half-metallicity, of interest for spintronic applications.\cite{He}

The magnetocaloric effect (MCE) in the FM vdW materials gives additional insight into the magnetic properties. Bulk CrI$_3$ exhibits anisotropic $-\Delta S_M^{max}$ with the values of 4.24 and 2.68 J kg$^{-1}$ K$^{-1}$ at 5 T for out-of-plane and in-plane fields, respectively,\cite{YuLIU} however little is known about VI$_3$.

In the present work we focus on the nature of the FM transition in bulk VI$_3$ single crystals. We have investigated the critical behavior by the modified Arrott plot and a critical isotherm analysis, whilst the magnetocaloric effect was also studied by heat capacity and magnetization measurements near $T_c$. Critical exponents $\beta$ = 0.244(5) with $T_c$ = 50.10(2) K, $\gamma$ = 1.028(12) with $T_c$ = 49.97(5) K, and $\delta$ = 5.24(2) at $T_c$ = 50 K, suggest that the magnetic transition in VI$_3$ is of second-order and that it is situated near a critical point from three- to two-dimensional. This is further confirmed by the scaling analysis of magnetic entropy change $-\Delta S_M(T,H)$, in which the rescaled $-\Delta S_M(T,H)$ collapse on a universal curve independent on temperature and field.

\section{Experimental details}

Bulk VI$_3$ single crystals were fabricated by the chemical vapor transport method starting from an intimate mixture of vanadium powder (99.95 $\%$, Alfa Aesar) and anhydrous iodine beads (99.99 $\%$, Alfa Aesar) with a molar ratio of 1 : 3. The starting materials were sealed in an evacuated quartz tube, placed inside a multi-zone furnace and then reacted over a period of 7 days with the source zone at 650 $^\circ$C, the middle growth zone at 550 $^\circ$C, and the third zone at 600 $^\circ$C. The crystal structure was characterized by powder x-ray diffraction (XRD) in the transmission mode at 28-D-1 beamline of the National Synchrotron Light Source II (NSLS II) at Brookhaven National Laboratory (BNL). Data were collected using a 0.5 mm$^2$ beam with wavelength $\lambda \sim$ 0.1668 {\AA}. A Perkin Elmer 2D detector (200 $\times$ 200 microns) was placed orthogonal to the beam path 990 mm away from the sample. The single crystal XRD were taken with Cu K$_{\alpha}$ ($\lambda=0.15418$ nm) radiation of Rigaku Miniflex powder diffractometer. The element analysis was performed using an energy-dispersive x-ray spectroscopy (EDS) in a JEOL LSM-6500 scanning electron microscope, confirming a stoichiometric VI$_3$ single crystal. The magnetization data as a function of temperature and field were collected using Quantum Design MPMS-XL5 system. The heat capacity was measured in Quantum Design PPMS-9 system.

\section{RESULTS AND DISCUSSIONS}

\subsection{Structure and basic magnetic properties}

\begin{figure}
\centerline{\includegraphics[scale=1]{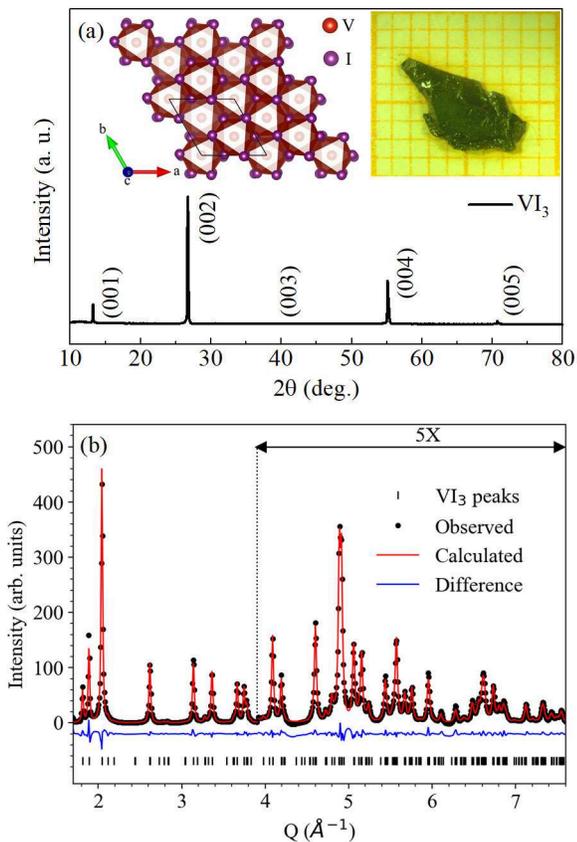}}
\caption{(Color online) (a) Single crystal x-ray diffraction (XRD) pattern of VI$_3$. Inset shows the $ab$ plane structure and representative single crystal. (b) Refinement of synchrotron powder XRD data of VI$_3$ at room temperature.}
\label{XRD}
\end{figure}

\begin{figure}
\centerline{\includegraphics[scale=1]{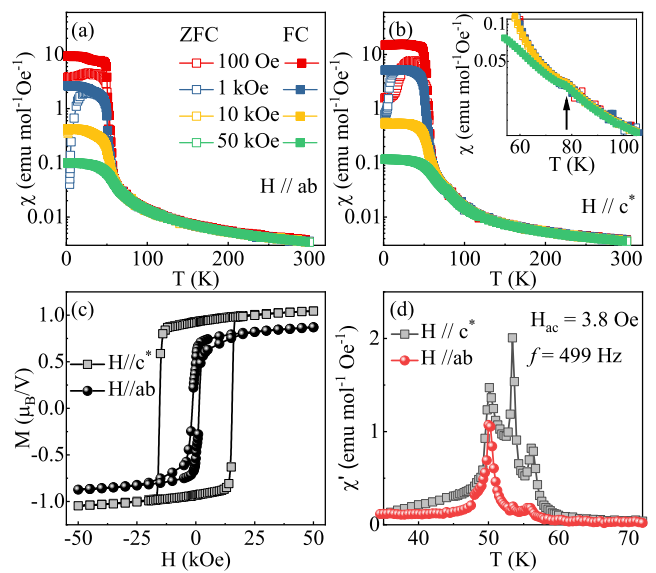}}
\caption{(Color online) Temperature dependence of dc magnetic susceptibility $\chi$ for VI$_3$ measured in various fields applied (a) in the $ab$ plane and (b) along the $c^*$ axis, respectively. (c) Field dependence of magnetization measured at $T$ = 2 K. (d) Ac susceptibility real part $\chi^\prime(T)$ as a function of temperature measured with oscillated ac field of 3.8 Oe and frequency of 499 Hz applied in the $ab$ plane and along the $c^*$ axis.}
\label{MTH}
\end{figure}

The as-grown single crystals are shiny black platelets with lateral dimensions up to several millimeters. In the single-crystal XRD scan [Fig. 1(a)], only $(00l)$ peaks are detected, indicating that the plate-shaped surface parallel to the $ab$ plane, and we assign the axis $c^*$ is normal to the plane. The layer spacing of VI$_3$ is calculated as 6.67(1) {\AA}, close to the reported value.\cite{Son,Kong,Tian} Rietveld powder diffraction analysis was carried out on data obtained from the raw 2D diffraction data integrated and converted to intensity versus $Q$ using the Fit2d software where $Q = 4\pi sin\theta / \lambda$ is the magnitude of the scattering vector.\cite{Hammersley} The refinement was performed using GSAS-II modeling suite.\cite{Toby} Figure 1(b) shows the refinement result of synchrotron powder XRD data of VI$_3$ at room temperature (space group $R\overline{3}$). The determined lattice parameters are $a$ = 6.9137(11) {\AA} and $c$ = 19.9023(21) {\AA}.

Figures 2(a) and 2(b) present the temperature dependence of dc magnetic susceptibility measured in the fields ranging from 100 Oe to 50 kOe applied in the $ab$ plane and along the $c^*$ axis, respectively. It is clearly seen that VI$_3$ exhibits a ferromagnetic transition near $T_c$ = 50 K for both magnetic field directions, consistent with the previous reports.\cite{Son,Kong,Tian} The magnetic susceptibility is nearly isotropic in H = 50 kOe, however, significant magnetic anisotropy is observed in low fields. When T $<$ T$_c$, the divergence of zero-field cooling (ZFC) and field-cooling (FC) curves exhibit a characteristic behavior of possible spin-glass state with the temperature of divergence decreasing with increasing field. Besides this, the magnetic domain creep, i.e., the magnetic domain walls jump from one pinning site to another, can also lead to this kind of irreversible behavior.\cite{Tian} The evolution of ferromagnetic domain as a function of magnetic field and temperature was further investigated,\cite{Kong} confirming the ferromagnetism and a small domain-wall-energy in VI$_3$. It should be noted that there is an additional weak anomaly at 80 K for $\mathbf{H\parallel c^*}$ [inset in Fig. 2(b)], which is field-independent. A structural phase transition accompanies similar feature in the susceptibility of CrI$_3$,\cite{McGuire} indicating strong spin-lattice coupling.

Isothermal magnetization at $T$ = 2 K [Fig. 2(c)] shows saturation moments of $M_s \approx$ 0.72 $\mu_B$/V and 0.95 $\mu_B$/V for $\mathbf{H\parallel ab}$ and $\mathbf{H\parallel c^*}$, respectively. The value is smaller than the expected saturated moment of 2 $\mu_B$ for V$^{3+}$ ion. The difference of saturation magnetization for the two directions is also unusual, which may be due to anisotropic $g$ factor with unquenched orbital angular moment, calling for further neutron scattering and/or electron spin resonance studies.\cite{Son,Kong,Tian} The coercive field is about 15 kOe for $\mathbf{H\parallel c^*}$, much larger than that of 1.5 kOe for $\mathbf{H\parallel ab}$, suggesting a hard ferromagnet behavior and the easy $c^*$ axis. The coercive field is significantly larger than that in CrI$_3$ with fully filled Cr$^{3+}$ orbitals. Son et al. proposed that the smaller saturated moment in V$^{3+}$ driven by the smaller number of $d$-orbital spin and the larger magnetic anisotropy coming from the partially filled $t_{2g}$ $d$-band of V$^{3+}$ would lead to the larger coercive field in VI$_3$ when compared with CrI$_3$.\cite{Son} Ac susceptibility was further measured with zero field cooling at oscillated ac field of 3.8 Oe and frequency of 499 Hz. Three distinct peaks in the real part $\chi^\prime(T)$ along the $c^*$ axis [Fig. 2(d)], one strong peak for both directions corresponding the PM-FM transition at $T_c$ = 50 K and two additional peaks above $T_c$, as well as the weak anomalies at the same temperatures in the $ab$ plane, indicating a complex multiple-step magnetic ordering in VI$_3$.

\begin{figure}
\centerline{\includegraphics[scale=1]{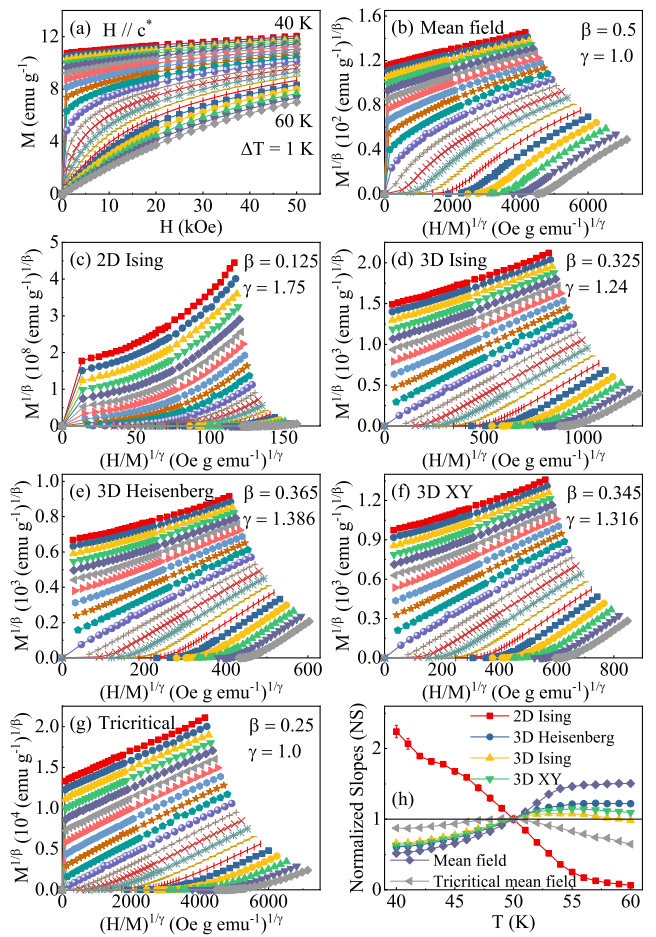}}
\caption{(Color online) (a) Typical initial isothermal magnetization curves measured in out-of-plane fields from 40 to 60 K with a temperature step of 1 K for VI$_3$. (b) The Arrott plot of $M^2$ vs $H/M$. The $M^{1/\beta}$ vs $(H/M)^{1/\gamma}$ plot with parameters of (c) 2D Ising model, (d) 3D Ising model, (e) 3D Heisenberg model, (f) 3D XY model, and (g) Tricritical mean-field model. (h) Temperature dependence of the normalized slopes $NS = S(T)/S(T_c)$ for different models.}
\label{Arrot}
\end{figure}

\begin{figure}
\centerline{\includegraphics[scale=1]{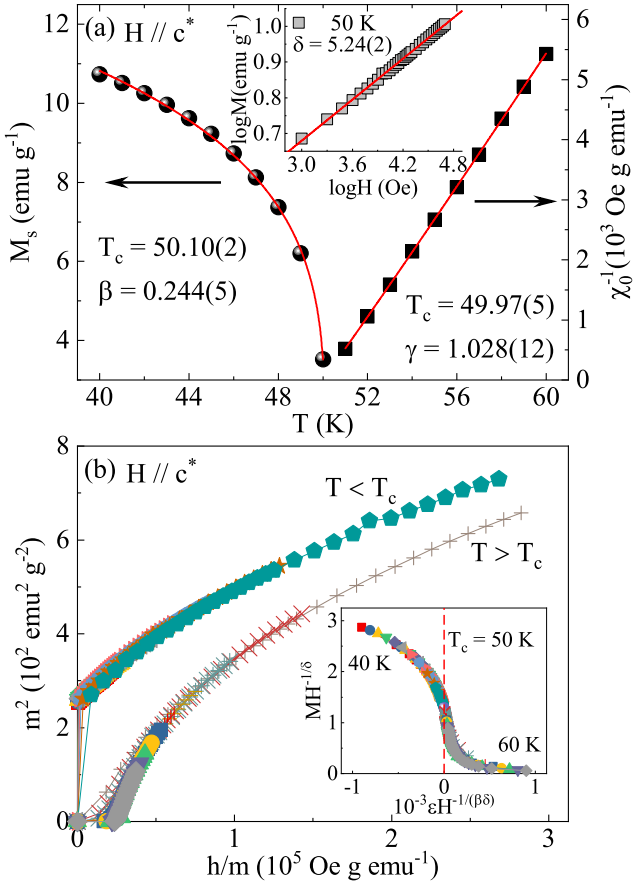}}
\caption{(Color online) (a) Temperature dependence of the spontaneous magnetization $M_s$ (left) and the inverse initial susceptibility $\chi_0^{-1}$ (right) with solid fitting curves. Inset shows the log$M$ vs log$H$ collected at 50 K with linear fitting curve. (b) Scaling plots of $m^2$ vs $h/m$ with the scaled magnetization $m\equiv\varepsilon^{-\beta}M(H,\varepsilon)$ and the scaled field $h\equiv\varepsilon^{-(\beta+\gamma)}H$ below and above $T_c$ with critical exponents $\beta = 0.244$, $\gamma = 1.028$, and $\delta = 5.24$ for VI$_3$. Inset shows the rescaling of the $M(H)$ curves by $MH^{-1/\delta}$ vs $\varepsilon H^{-1/(\beta\delta)}$.}
\label{KF}
\end{figure}

\subsection{Critical behavior}

To determine the accurate $T_c$, we first considered the well-known Arrott plot.\cite{Arrott1} Magnetization isotherms along the easy $c$ axis were measured in the vicinity of $T_c$ [Fig. 3(a)]. The Arrott plot involves the mean-field critical exponents $\beta$ = 0.5 and $\gamma$ = 1.0.\cite{Arrott1} Based on this, magnetization isotherms $M^2$ vs $H/M$ should be a set of parallel straight lines and the isotherm at $T_c$ should pass through the origin. As is seen, all curves in the Arrott plot of VI$_3$ are nonlinear [Fig. 3(b)], with a downward curvature, demonstrating that the mean-field model does not work for VI$_3$. Based on Banerjee$^\prime$s criterion,\cite{Banerjee} we can estimate the order of the magnetic transition through the slope of the straight line. First (second) order phase transition corresponds to negative (positive) slope. Therefore, the downward slope reveals a second-order PM-FM transition in VI$_3$.

In the vicinity of $T_c$ the second order phase transition is governed by magnetic equation of state and is characterized by critical exponents $\beta$, $\gamma$ and $\delta$ that are mutually related.\cite{Stanley} Spontaneous  magnetization $M_s$ and inverse initial susceptibility $\chi_0^{-1}$, below and above $T_c$ can be used to obtain $\beta$ and $\gamma$ whereas $\delta$ is the critical isotherm exponent. Hence, from magnetization:
\begin{equation}
M_s (T) = M_0(-\varepsilon)^\beta, \varepsilon < 0, T < T_c,
\end{equation}
\begin{equation}
\chi_0^{-1} (T) = (h_0/m_0)\varepsilon^\gamma, \varepsilon > 0, T > T_c,
\end{equation}
\begin{equation}
M = DH^{1/\delta}, T = T_c,
\end{equation}
where $\varepsilon = (T-T_c)/T_c$ is the reduced temperature, and $M_0$, $h_0/m_0$ and $D$ are the critical amplitudes.\cite{Fisher} For the original Arrott plot, $\beta$ = 0.5 and $\gamma$ = 1.0.\cite{Arrott1} In a more general case, the Arrott-Noaks equation of state provides modification of Arrott plot:\cite{Arrott2}
\begin{equation}
(H/M)^{1/\gamma} = a\varepsilon+bM^{1/\beta},
\end{equation}
where $\varepsilon = (T-T_c)/T_c$ and $a$ and $b$ are fitting constants. Since the mean-field model does not work, we adopt the modified Arrott plot in order to better understand the nature of the PM-FM transition in VI$_3$.

Figures 3(c)-3(g) exhibit the modified Arrott plots using possible exponents from 2D Ising ($\beta = 0.125, \gamma = 1.75$), 3D Ising ($\beta = 0.325, \gamma = 1.24$), 3D Heisenberg ($\beta = 0.365, \gamma = 1.386$), 3D XY ($\beta = 0.345, \gamma = 1.316$), and tricritical mean-field ($\beta = 0.25, \gamma = 1.0$) models.\cite{Kaul,Khuang,LeGuillou} The modified Arrott plot should be a set of parallel lines in the high field region with the same slope $S(T) = dM^{1/\beta}/d(H/M)^{1/\gamma}$. The model which fits the data best is selected via the normalized slope [$NS = S(T)/S(T_c)$] that compares with the ideal value of unity. Plot of $NS$ vs $T$ for different models is also presented in Fig. 3(h). It is clearly seen that the $NS$ of 2D Ising model shows the largest deviation from unity. The $NS$ of 3D Ising model is close to $NS = 1$ mostly above $T_c$, while that of tricritical mean field model is the best below $T_c$.

Following the methods of Pramanik and Banerjee,\cite{Pramanik} the linearly extrapolated $M_s$ and $H/M$ are plotted as a function of temperature in Fig. 4(a). The solid lines are fitted lines according to Eqs. (1) and (2). The critical exponents $\beta = 0.244(5)$, with $T_c = 50.10(2)$ K, and $\gamma = 1.028(12)$, with $T_c = 49.97(5)$ K, are obtained. As we can see, the value of $\gamma$ is close to that of tricritical mean-field model ($\gamma = 1.0$), while $\beta$ lies between the values of tricritical mean-field ($\beta = 0.25$) and 2D XY model ($\beta = 0.23$).\cite{Bramwell} It is summarized that the value of $\beta$ for a 2D magnet should be within a window $0.1 \leq \beta \leq 0.25$.\cite{Taroni} Therefore, the obtained critical exponents suggest that the magnetic transition of VI$_3$ is situated close to a three- to two-dimensional critical point, in contrast to those of CrI$_3$ exhibiting 3D critical behavior and Cr$_2$(Si,Ge)$_2$Te$_6$ showing 2D Ising-type coupled with a long-range interaction.\cite{YuL,GT,BJLIU,YULIU,GTLIN,JC} According to Eq. (3), the $M(H)$ at $T_c$ should be a straight line in log-log scale with the slope of $1/\delta$. Such fitting yields $\delta = 5.24(2)$ [inset in Fig. 4(a)]. The Widom relation gives $\delta = 1+\gamma/\beta$.\cite{Widom} From $\beta$ and $\gamma$ obtained with the modified Arrott plot, $\delta$ is calculated to be 5.21(4), which is agree with that obtained from critical isotherm analysis.

Scaling analysis can be used to estimate the reliability of the obtained critical exponents. Near phase transition the magnetic equation of state is:
\begin{equation}
M(H,\varepsilon) = \varepsilon^\beta f_\pm(H/\varepsilon^{\beta+\gamma}),
\end{equation}
where $f_+$ for $T > T_c$ and $f_-$ for $T < T_c$, respectively, are the regular functions. Eq.(5) can be expressed via rescaled magnetization $m\equiv\varepsilon^{-\beta}M(H,\varepsilon)$ and rescaled field $h\equiv\varepsilon^{-(\beta+\gamma)}H$ as
\begin{equation}
m = f_\pm(h).
\end{equation}
For the correct scaling relations and correct choice of $\beta$, $\gamma$, and $\delta$, scaled $m$ and $h$ fall on universal curves above $T_c$ and below $T_c$, respectively. Figure 4(b) presents the scaled $m^2$ vs $h/m$ that collapse on two separate branches below and above $T_c$, respectively, confirming proper treatment of the critical regime. The scaling equation of state also takes another form
\begin{equation}
\frac{H}{M^\delta} = k(\frac{\varepsilon}{H^{1/\beta}}),
\end{equation}
where $k(x)$ is the scaling function. From Eq. (7), all the experimental data should fall into a single curve. This is indeed seen in the inset of Fig. 4(b); the $MH^{-1/\delta}$ vs $\varepsilon H^{-1/(\beta\delta)}$ experimental data collapse into a single curve and the $T_c$ is located at the zero point of the horizontal axis.

\subsection{Magnetic entropy change}

\begin{figure}
\centerline{\includegraphics[scale=1]{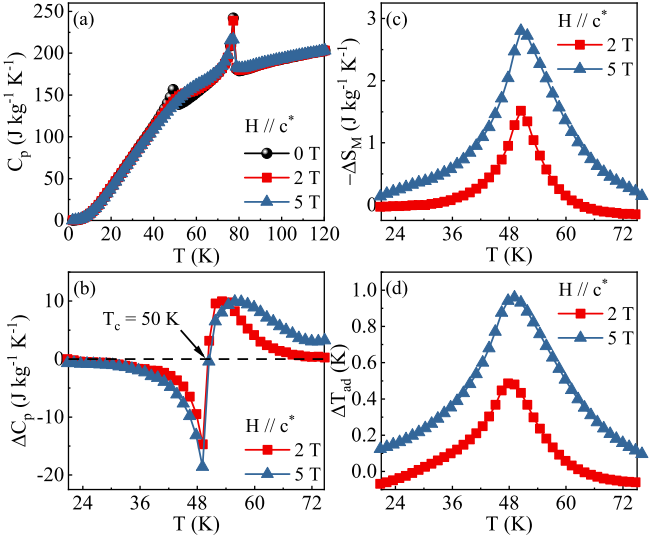}}
\caption{(Color online) Temperature dependence of (a) the specific heat $C_p$, (b) the specific heat change $\Delta C_p$, (c) the magnetic entropy change $-\Delta S_M$, and (d) the adiabatic temperature change $\Delta T_{ad}$ for VI$_3$ at the indicated out-of-plane fields.}
\label{heatcapacity}
\end{figure}

\begin{figure}
\centerline{\includegraphics[scale=1]{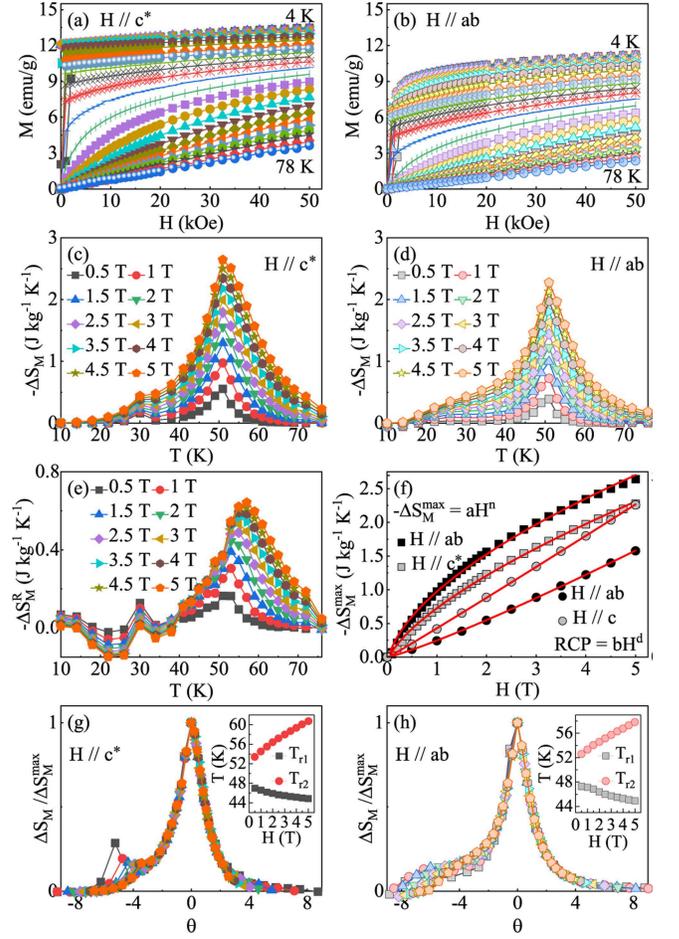}}
\caption{(Color online) Initial isothermal magnetization curves measured in (a) $\mathbf{H\parallel c^*}$ and (b) $\mathbf{H\parallel ab}$ with a temperature step of 2 K. The magnetic entropy change $-\Delta S_M$ obtained from magnetization at indicated field changes with (c) $\mathbf{H\parallel c^*}$ and (d) $\mathbf{H\parallel ab}$, respectively. (e) Temperature dependence of $-\Delta S_M^R$ obtained by rotating from the $ab$ plane to the $c$ axis in various fields. (f) Field dependence of the maximum magnetic entropy change $-\Delta S_M^{max}$ (left) and the relative cooling power RCP (right) with power law fitting in red solid lines. The normalized $\Delta S_M$ as a function of the rescaled temperature $\theta$ for (g) $\mathbf{H\parallel c^*}$ and (h) $\mathbf{H\parallel ab}$, respectively.}
\label{magenticentropychange}
\end{figure}

Figure 5(a) shows the temperature dependence of heat capacity $C_p$ at different fields. A sharp peak at $T \sim$ 80 K is observed. There is almost no shift when the magnetic field changes, corresponding to the structural transition, consistent with the susceptibility anomaly [inset in Fig. 2(b)]. In contrast, the peak of magnetic order at lower temperature is gradually suppressed when the magnetic field increases. At $T_c \sim$ 50 K, the heat capacity change $\Delta C_p = C_p(T,H)-C_p(T,0)$ exhibits a sharp change from negative to positive [Fig. 5(b)]. The entropy $S(T,H)$ = $\int_0^T C_p(T,H)/T dT$ and the magnetic entropy change $-\Delta S_M(T,H) = S_M(T,H)-S_M(T,0)$. The adiabatic temperature change $\Delta T_{ad}$ caused by the field change can be obtained by $\Delta T_{ad}(T,H) = T(S,H)-T(S,0)$, where $T(S,H)$ and $T(S,0)$ are the temperatures in $H \neq 0$ and $H = 0$, respectively, at constant total entropy $S(T,H)$. Figures 6(b) and 6(c) show the temperature dependence of $-\Delta S_M$ and $\Delta T_{ad}$ estimated from heat capacity with out-of-plane field change. The maxima of $-\Delta S_M$ and $\Delta T_{ad}$ increase with increasing field and reach the values of 2.80 J kg$^{-1}$ K$^{-1}$ and 0.96 K, respectively, with the field change of 5 T. The obtained $-\Delta S_M$ and $\Delta T_{ad}$ of VI$_3$ are significantly smaller than those of well-known magnetic refrigerating materials, such as Gd$_5$Si$_2$Ge$_2$, LaF$_{13-x}$Si$_x$, and MnP$_{1-x}$Si$_x$,\cite{GschneidnerJr} however, comparable with those of Cr(Br,I)$_3$ and Cr$_2$(Si,Ge)$_2$Te$_6$.\cite{Xiaoyun,YuLIU,YL}

Figures 6(a) and 6(b) show the initial isothermal magnetization with the temperature ranging from 4 K to 78 K for $\mathbf{H\parallel c^*}$ and $\mathbf{H\parallel ab}$, respectively. The magnetic entropy change
\begin{equation}
\Delta S_M(T,H) = \int_0^H \left(\frac{\partial S}{\partial H}\right)_TdH = \int_0^H \left(\frac{\partial M}{\partial T}\right)_HdH,
\end{equation}
where $\left(\frac{\partial S}{\partial H}\right)_T$ = $\left(\frac{\partial M}{\partial T}\right)_H$ is based on Maxwell's relation.\cite{Amaral} For magnetization measured at small (H,T) intervals,
\begin{equation}
\Delta S_M(T_i,H) = \frac{\int_0^HM(T_i,H)dH-\int_0^HM(T_{i+1},H)dH}{T_i-T_{i+1}}.
\end{equation}
Figures 6(c) and 6(d) give the calculated $-\Delta S_M(T,H)$ as a function of temperature in $\mathbf{H\parallel c^*}$ and $\mathbf{H\parallel ab}$, respectively. All the $-\Delta S_M$ curves exhibit a pronounced peak at $T_c$. The maxima $-\Delta S_M$ reach 2.64 and 2.27 J kg$^{-1}$ K$^{-1}$ with out-of-plane and in-plane field change of 5 T, respectively. In view of a large magnetic anisotropy in VI$_3$, the rotational magnetic entropy change $\Delta S_M^R$ is calculated as $\Delta S_M^R(T,H) = \Delta S_M(T,H_c)-\Delta S_M(T,H_{ab})$. Figure 6(e) shows the temperature-dependent $-\Delta S_M^R$ of VI$_3$, which is smaller than that of CrI$_3$.\cite{YuLIU}

The magnetic entropy change is also correlated with the intrinsic magnetic coupling through a series of critical exponents. The maximal magnetic entropy change $-\Delta S_M^{max} = aH^n$.\cite{VFranco,VFranco1} The relative cooling power $RCP$ is defined as $RCP = -\Delta S_M^{max} \times \delta T_{FWHM}$, where $\delta T_{FWHM}$ is the full-width at half maximum, and $RCP = bH^d$.\cite{VFranco,VFranco1} Figure 6(f) presents the field dependence of $-\Delta S_M^{max}$ and RCP. Fitting of $-\Delta S_M^{max}$ and RCP give that $n = 0.58(2)$ and $d = 1.02(1)$ for out-of-plane field, while $n = 0.67(1)$ and $d = 1.15(1)$ for in-plane field. As is known, the exponents $n$ and $d$ are correlated with the critical exponents as $n=1+(\beta-1)/(\beta+\gamma)$ and $d=1+1/\delta$.\cite{Franco} The obtained $n$ is close to that of 3D Ising model ($n$ = 0.569) for out-of-plane field and approaches the value of mean-field model ($n = 0.667$) for in-plane field.

The $-\Delta S_M$ scaling analysis is assessed from normalizing all the $-\Delta S_M$ curves against their maxima $-\Delta S_M^{max}$, i.e., $\Delta S_M/\Delta S_M^{max}$ by temperature $\theta$ rescaling based on:\cite{Franco}
\begin{equation}
\theta_- = (T_{peak}-T)/(T_{r1}-T_{peak}), T<T_{peak},
\end{equation}
\begin{equation}
\theta_+ = (T-T_{peak})/(T_{r2}-T_{peak}), T>T_{peak},
\end{equation}
where $T_{r1}$ and $T_{r2}$ are the temperatures of the two reference points that have been selected as those corresponding to $\Delta S_M(T_{r1},T_{r2}) = \Delta S_M^{max}/2$. It could be seen that the $-\Delta S_M(T,H)$ in different magnetic fields fall on a single line near $T_c$ [Figs. 6(g) and 6(h)]. The well scaling of $-\Delta S_M(T,H)$ curves near $T_c$ indicate that the magnetic phase transition of VI$_3$ is of second-order. The slight deviation at low temperature is most likely contributed by its magnetic anisotropy effect.

\section{CONCLUSIONS}

In summary, we have studied the critical behavior and magnetocaloric effect around the FM-PM transition in VI$_3$ single crystal. The PM-FM transition in VI$_3$ is identified to be of second order. The critical exponents $\beta$, $\gamma$, and $\delta$ suggest the ferromagnetic phase transition in VI$_3$ is situated close to a 3D to 2D critical point. Considering its ferromagnetism can be maintained upon exfoliating bulk crystals down to a single layer, further investigation on the size-dependent properties is of interest.

\textit{Note added}. We became aware of several related works after the completion of our work.\cite{JYan,Elena,Dolezal}

\section*{Acknowledgements}

This work was funded by the Computation Material Science Program (Y.L. and C.P.). This research used the 28-ID-1 beamline of the National Synchrotron Light Source II, a U.S. DOE Office of Science User Facility operated for the DOE Office of Science by Brookhaven National Laboratory under Contract No. DE-SC0012704.

\end{document}